\documentstyle[12pt]{article}
\begin{document}
\begin{center}
{\huge\bf Onsets of avalanches in the BTW model}
\end{center}
\vspace {0.4 cm}
\begin{center} {\large Ajanta Bhowal$^*$} \end{center}
\vspace {0.3 cm}
\begin{center} {\it Institute for Theoretical Physics} \end{center}
\begin{center} {\it University of Cologne, D-50923 Cologne, 
Germany} \end{center}
\vspace {1.0 cm}
\noindent {\bf Abstract:} 
The onsets of toppling and dissipation in the BTW model are studied by
computer simulation. The distributions of these two onset times and their
dependences on the system size are also studied. Simple power law
dependences of these two times on the system size are observed and
the exponents are estimated. The fluctuation
of the average (spatial) height in the subcritical region is studied 
and observed to increase very rapidly near the SOC point.

\vspace {3 cm}
\leftline {\bf Keywords: BTW model, SOC}
\leftline {\bf PACS Numbers: 05.50 +q}
\vspace {6 cm}
\leftline {----------------------------}
\leftline {\bf $^*$E-mail:ajanta@hp1.saha.ernet.in}
\newpage

\leftline {\bf I. Introduction}

There exists some extended driven dissipative systems in nature which
show self-organised criticality (SOC). This phenomena of SOC is
characterised by sponteneous evolution into a steady state which shows
long-range spatial and temporal correlations. The concept of SOC was
introduced by Bak et al in terms of a simple cellular automata model [1].
The steady state dynamics of the model shows a power law in the probability
distributions for the occurence of the relaxation (avalanches) clusters of a
certain size, area, lifetime, etc. Extensive work has been done so far to
study the properties of the model in the steady SOC state[2-10]. 
Using the commutative property of the particle addition operator this
model has been solved exactly [2]. Several properties of this critical state,
e.g., entropy, height correlation, height probabilities, etc have been 
calculated analytically [2-3]. But the critical exponents have not been 
calculated analytically and hence an extensive numerical efforts have
been performed to  estimate various exponents [4-7]. The values of critical
exponent for size and lifetime distributions of avalanches starting at
the boundary have been calculated [8]. Recently, the avalanche exponents
were estimated using the renormalization scheme [9].

Very few efforts are made to study the systematic evolutions of the system
 towards the SOC state. 
In the subcritical region the response of a pulsed addition of particles
has been studied and it has been observed [11] that the  ratio of response
time ($\delta t$) to the perturbation time ($\Delta t$) diverges as 
the system approaches the critical state 
(i.e., $R={\delta t \over \Delta t} \sim (z_c -z)^{-\gamma}$), 
which implies the critical slowing down in this model.
Grassberger and Manna [12] also anticipated such kind of scaling
behaviour ($<s>, <t> \sim (z_c -z)^{-\beta}$) and obtained only for 
two dimension. 

In this paper, we have focussed on the subcritical region of the time
evolution of the BTW model. We have studied how
the onset time of toppling and that of dissipation (escape of particle through
the boundary) vary with the system size. The distributions of these two 
onset times have also been found.
We have also studied the growth of fluctuations of the height variable (averaged
over all sites) as the system attains the critical state.

\bigskip
\leftline {\bf II. The model and simulation}

The BTW model is a lattice automata model which shows some important
properties of the dynamics of the system which evolves spontaneously into
a critical state. We consider a two dimensional square lattice of size
$L \times L$. The description of the model is the following: At each site
$(i,j)$
of the lattice a variable (so called height) $z(i,j)$ is associated
which can take positive integer values. In every time step, one particle
is added to a randomly chosen site according to
\begin{equation}
z(i,j)=z(i,j)+1. 
\end{equation}

\noindent If, at any site the height variable exceeds a critical value
$z_m$ (i.e., if $z(i,j) \geq z_m$), then that site becomes unstable and
it relaxes by a toppling. As an unstable site
topples, the value of the height variable, of that site is decreased by 4
units and that, of each of the four of its neighbouring sites  increased   
by unity (local conservation), i.e.,
\begin{equation}
z(i,j)=z(i,j)-4
\end{equation}
\begin{equation}
z(i, j \pm 1) = z(i , j \pm 1) +1 ~~{\rm and}~~
z(i \pm 1,j) = z(i\pm 1 ,j ) +1 
\end{equation}
\noindent for $z(i,j) \geq z_m$. Each boundary site is attached  to
an additional site which acts as a sink. We use here the 
open boundary conditions  so that the system can
dissipate  through the boundary. In our simulation, we have
taken $z_m = 4$. The main investigations in this paper can be divided as
follows:

(1) Studies regarding the onset times of the toppling and
dissipations, where
we allow the system to evolve under the BTW dynamics (following eqns 1-3)
starting from an initial condition with all the sites having $z=0$. With
the evolution of time, the height at different sites first increases due
to random addition of particles. As soon as the height at any site reaches
(or exceeds)
the maximum value ($z_m = 4$), that site topples. We call this time, when
the toppling starts in the system,  the onset time of toppling, $T^t_o$.
In most of the cases, the interior sites first topple and then, after some
time toppling occurs at the boundary sites.
The system starts to dissipate (through boundary) as soon as
any boundary site topples. The time, taken by the system to start
dissipation, is called  onset time of dissipation, $T^d_o$. 
The onset of toppling of the boundary site (to start
dissipation) can occur in either of the two ways: (i) via
the primary avalanche of the boundary site, (ii) through the secondary
avalanche followed by a primary avalanche, initiated at any interior
site of the lattice.  These two 
onset times of the system may change appreciably as one change the 
random sequence of addition of particles.
As a result, these two times may have wide fluctuations
in their distributions. We have studied here the statistical distributions
of these two times, $T^t_o$ and $T^d_o$. The size (of the system) 
dependences of these
two onset times  are also studied.

(2) Studies regarding the growth of fluctuations near the SOC point:
The average (spatial) value of $z$, i.e.,
$$\bar z= (1/N)\sum_{i=1}^N z_i ~~~~~~~~~~~~~(N=L^2)$$
\noindent increases almost linearly with the time in the subcritical
 region and then in the critical state, 
$\bar z$ becomes steady apart from some fluctuations [4]. Grassberger
and Manna [12] have studied the system size dependence of the fluctuation 
of $\bar z$ in the critical state.  
Here, we pay  some attention to study how the fluctuation of $\bar z$
 changes as the system approaches SOC for a particular length ($L=100$).
Here, the fluctuation of $\bar z$ means the standard deviation 
of $\bar z$, at a particular time, for different random sample, i.e.,
$$\delta z= \sqrt {{1\over N_s} \sum_{l=1}^{N_s}(\bar z_l -<\bar z > )^2 }$$

\noindent where $\bar z_l$ is the value of $\bar z$ for $l^{th}$ random sample
 and 
$ <\bar z>$ is the average value of $\bar z$ calculated from $N_s$ number of
different random samples, i.e.,
$$<\bar z >= (1/N_s)\sum_{l=1}^{N_s} z_l  .$$

\bigskip
\leftline {\bf III. Results}

In our simulation, for a fixed system size ($L = 100$), the distribution
of $T^t_o$ and $T^d_o$ are obtained
from $10^4$  different samples. 
Figure 1 shows the distribution of the 
onset time of toppling ($D(T^t_o)$) and the distribution of the
onset time of dissipation ($D(T^d_o)$). It has been observed that the onset
times for toppling and that for dissipation vary in a wide range but they
have well defined symmetric distributions. The width of these two distributions
differ appreciably, where it is much much larger in the case of dissipations.
Since the number of boundary sites are smaller than that of the interior
sites the width of the distribution of $T_o^d$ is much larger than that
of $T_o^t$.

The onset time (average) for toppling and that for dissipation will depend 
upon the linear size ($L$) of the system.
The log-log plot of these two onset times  are depicted in Fig. 2.
The simulation results show that these dependences are  power law type.
The exponents are also estimated within limited accuracy.
$T^t_o \propto L^{a}$ and $T^d_o \propto L^{b}$,
 where $a \sim 1.48$ and $b \sim 1.70$.
All these data are obtained by averaging over 100 different
samples for $30\leq  L\leq 300$.

The growth of fluctuation of $\bar z$ is plotted against the time of evolution
of the system in the subcritical region in Fig. 3. It shows that the
fluctuation increases very sharply near the critical point (SOC).
The fluctuations of $\bar z$ are calculated for $L = 100$ using 400 random 
different samples.

\bigskip
\leftline {\bf IV. Summary}

We studied here, when the toppling and the dissipation start in the BTW model.
There is a well defined symmetric distribution
of the onset time for toppling and that for dissipation. These two
onset times depend (power law) on the system size with different 
exponents. 

>From Fig.1, we see that the maximum value of the onset time for
dissipation is of the order of $L^2$, when the average value of
the hight variable is of the order of unity (i.e., $\bar z \sim 1$).
Thus, there is a very little chance, that onset of toppling of the boundary 
site occurs due to the secondary avalanches of the interior site's 
avalanches. Almost  all the topplings of the boundary sites, for the
onset of dissipation, are due to the avalanches initiated at the boundary.
In this sense, onset of dissipation can be considered as onset of toppling
of the boundary site (due to the avalanche, initiated at that boundary
site). 

It is also important to know whether the system size
dependences of these two onset times  are  same
in the subcritical region for the other SOC 
models show the power law.

\bigskip

\leftline {\bf Acknowledgments:} The author is grateful to the Institute
for Theoretical Physics for
giving her the opportunity to use the computer at the University of Cologne,
Germany.

\newpage
\setlength{\unitlength}{0.240900pt}
\ifx\plotpoint\undefined\newsavebox{\plotpoint}\fi
\sbox{\plotpoint}{\rule[-0.200pt]{0.400pt}{0.400pt}}%


\bigskip

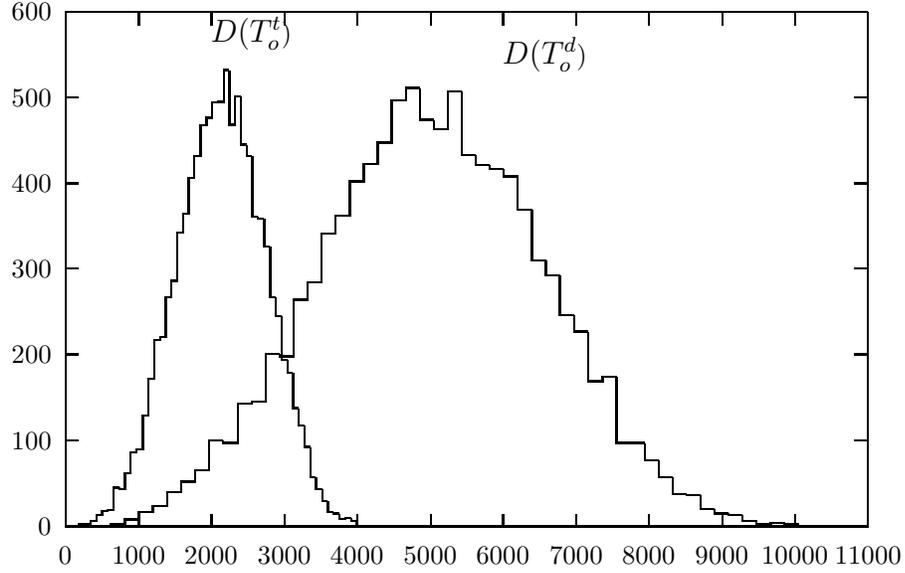
\noindent {\bf Fig. 1}. The unnormalized distributions of onset time
of toppling ($T^t_o$) and the onset time of dissipation ($T^d_o$).

\bigskip

\setlength{\unitlength}{0.240900pt}
\ifx\plotpoint\undefined\newsavebox{\plotpoint}\fi
\sbox{\plotpoint}{\rule[-0.200pt]{0.400pt}{0.400pt}}%
%
\bigskip

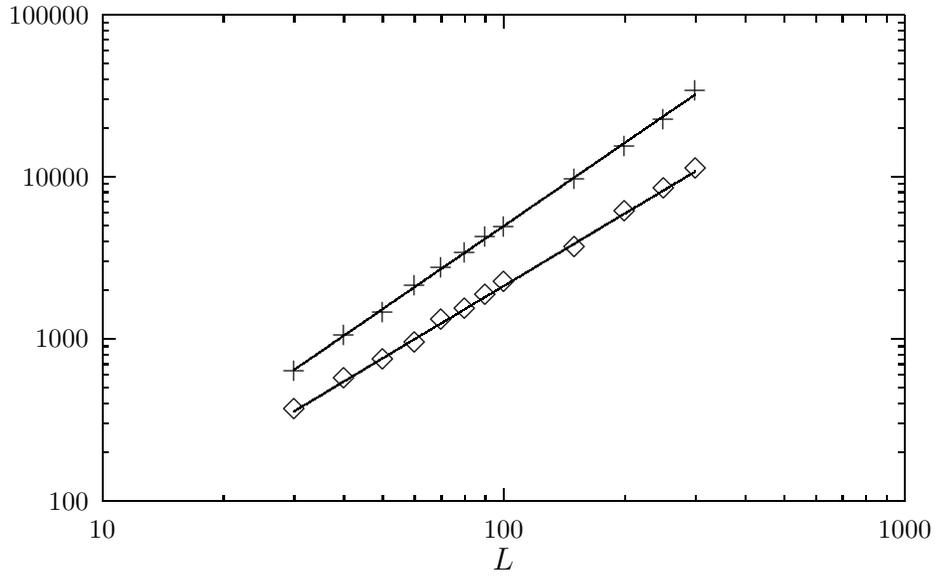
\noindent {\bf Fig. 2}. The onset times, for toppling ($T^t_o$) 
($\Diamond$) and dissipation ($T^d_o$)
($+$), are plotted against the system size $L$ in log-log scales. The solid
lines are linear best fit.
\newpage

\setlength{\unitlength}{0.240900pt}
\ifx\plotpoint\undefined\newsavebox{\plotpoint}\fi
\sbox{\plotpoint}{\rule[-0.200pt]{0.400pt}{0.400pt}}%
%
\bigskip

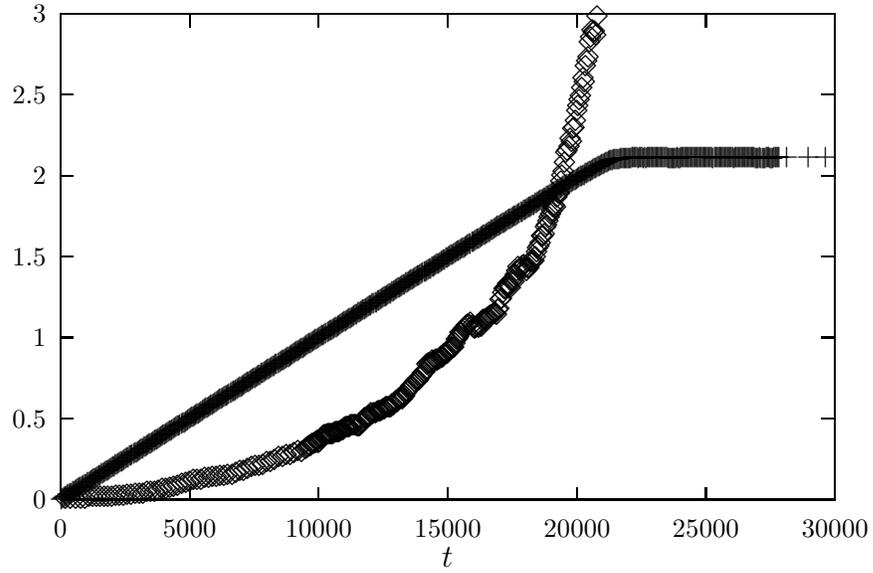
\noindent {\bf Fig. 3}. The time varations of $<z>$ ($+$) and 
$\delta z \times 10^3$ ($\Diamond$). At SOC $<z>$ = 2.124 (Ref. [4]).  


\vspace {0.6 cm}
\begin{thebibliography}{99}

\bibitem{btw} P. Bak, C. Tang and K. Wiesenfeld, Phys. Rev. Lett., {\bf 59}
(1987) 381; Phys. Rev. A, {\bf 38} (1988) 364.


\bibitem{dd} D. Dhar, Phys. Rev. Lett., {\bf 64} (1990) 1613;
S. N. Majumder and D. Dhar, J. Phys. A {\bf 24} (1991) L357.


\bibitem{evi} E. V. Ivashkevich, J. Phys. A {\bf 27} (1994) 3643;
V. B. Priezzhev, J. Stat. Phys, {\bf 74} (1994) 955.

\bibitem{ssm} S. S. Manna, J. Stat. Phys. {\bf 59} (1990) 509; Physica A
{\bf 179} (1991) 249

\bibitem{vbp} V. B. Priezzhev, D. V. Ktitarev and E. V. Ivashkevich,
Phys. Rev. Lett {\bf 76} (1996) 2093.

\bibitem{bb} A. Benhur and O. Biham, Phys. Rev. E {\bf 53} (1996) R1317.

\bibitem{lub} S. Lubek and K. D. Usadel, Phys. Rev. E {\bf 55} (1997) 4095.

\bibitem{ev} E. V. Ivashkevich, J. Phys. A {\bf 27 } (1994) L585.

\bibitem{piet} L. Pietronero, A. Vespignani and S. Zapperi, Phys. Rev. Lett.,
{\bf 72} (1994) 1690;
E. V. Ivashkevich, Phys. Rev. Lett, {\bf 76} (1996) 3368.

\bibitem{ab} A. Bhowal, Physica A {\bf 247} (1997) 327

\bibitem{ac} M. Acharyya and B. K. Chakrabarti, Physica A, {\bf 224} (1996)
254

\bibitem{gm} P. Grassberger and S. S. Manna, J. Phys. (Paris) {\bf 51} (1990)
1091


\end{thebibliography}
\end{document}